# Risk-Averse Power System Resilience Planning Under AI Data Center Demand Growth Using a Two-Stage DRO-CVaR MILP Optimization

Sharaf K. Magableh, Caisheng Wang, Oraib Dawaghreh, Xuesong Wang

*Department of Electrical and Computer Engineering*
Wayne State University, Detroit, MI, United States
{sharaf.magableh, cwang, oraib.dawaghreh, xswang}@wayne.edu

***Abstract*—The rapid growth of artificial intelligence-data centers is introducing significant demand variability and operational uncertainty into modern power systems, creating new challenges for resilience planning and grid operation. In addition to physical disruptions such as line outages and uncertain restoration processes, AI-driven loads can exhibit highly dynamic behavior. This paper presents a two-stage risk-averse Distributionally Robust Optimization (DRO)–Mixed Integer Linear Programming framework for enhancing distribution-system resilience under combined AI-driven demand uncertainty and physical disruptions. The proposed framework coordinates the strategic prepositioning and adaptive dispatch of flexible capacity modules, including battery energy storage systems and mobile diesel generation resources. A unified scenario representation captures AI-driven demand variability, line outages, and repair-time uncertainty. To improve robustness against uncertainty misspecification and extreme events, the model integrates Conditional Value-at-Risk within a DRO formulation. Validation on the IEEE 33-bus distribution system demonstrates reductions in Energy Not Supplied, improved served-load recovery, enhanced restoration performance, and increased system resilience under severe operating conditions.**

***Keywords*—AI Data Centers, Conditional Value-at-Risk, Distribution System Resilience, Flexible Capacity Modules.**

## I. INTRODUCTION

Modern power systems are entering a new era of stress and transformation. In addition to longstanding reliability challenges from extreme weather and natural disasters, the grid now faces unprecedented variability driven by artificial-intelligence (AI) data-center expansion, cloud computing, and large-scale electrification. These loads are large, rapidly changing, and highly volatile [1, 2], forcing planners and decision-makers to rethink how resilience is defined and managed.

The 2025 North American Electric Reliability Corporation (NERC) Report [3], identifies large, fast-changing electrical loads, as among the most pressing emerging reliability risks for the North American grid. These loads, particularly AI data centers and cryptocurrency operations, can exhibit rapid short-term fluctuations that challenge system balancing and stability. NERC also notes that a single incident in which 1.5 GW of load disconnected will almost instantaneously produce system-wide frequency excursions comparable to those following the sudden trip of a nuclear generating unit [3]. This report concludes that system designers must now treat these events as "*disruption equivalents*", which require improved modeling of both their magnitude and ramp-rate characteristics.

Globally, the International Energy Agency projects that electricity demand from data centers will more than double by 2030 to around 945 TWh. This is roughly equal to the entire consumption of Japan today [4]. In advanced economies, data centers are expected to account for more than 20% of the total load growth between now and 2030, putting an end to decades of flat or declining demand in mature markets. What's more, in the United States, nearly half of the projected electricity-demand growth through 2030 is expected to originate from data-center expansion [5]. These concentrations create localized stress on transmission and distribution networks, where multi-gigawatt facilities can swing demand by hundreds of megawatts within minutes during AI-training cycles [6]. U.S. data center load is projected to more than double by 2035, with aggregate capacity rising above 75 GW [7].

Collectively, these findings illustrate both the urgency and the opportunity to integrate AI-driven demand growth into resilience planning frameworks. Recent technical assessments and market studies highlighted that AI-intensive data centers can exhibit rapid short-term fluctuations, with measured ramping events reaching hundreds of megawatts within seconds [8]. Fig. 1 presents an AI/data-center demand-ramping event reported in [8], where the load decreases from approximately 450 MW to about 40 MW within seconds, remains at a reduced operating level for several hours, and then rapidly recovers. Such large and fast demand variations can introduce operational stresses comparable to those associated with major system disturbances, motivating their inclusion within resilience-oriented planning and optimization frameworks.

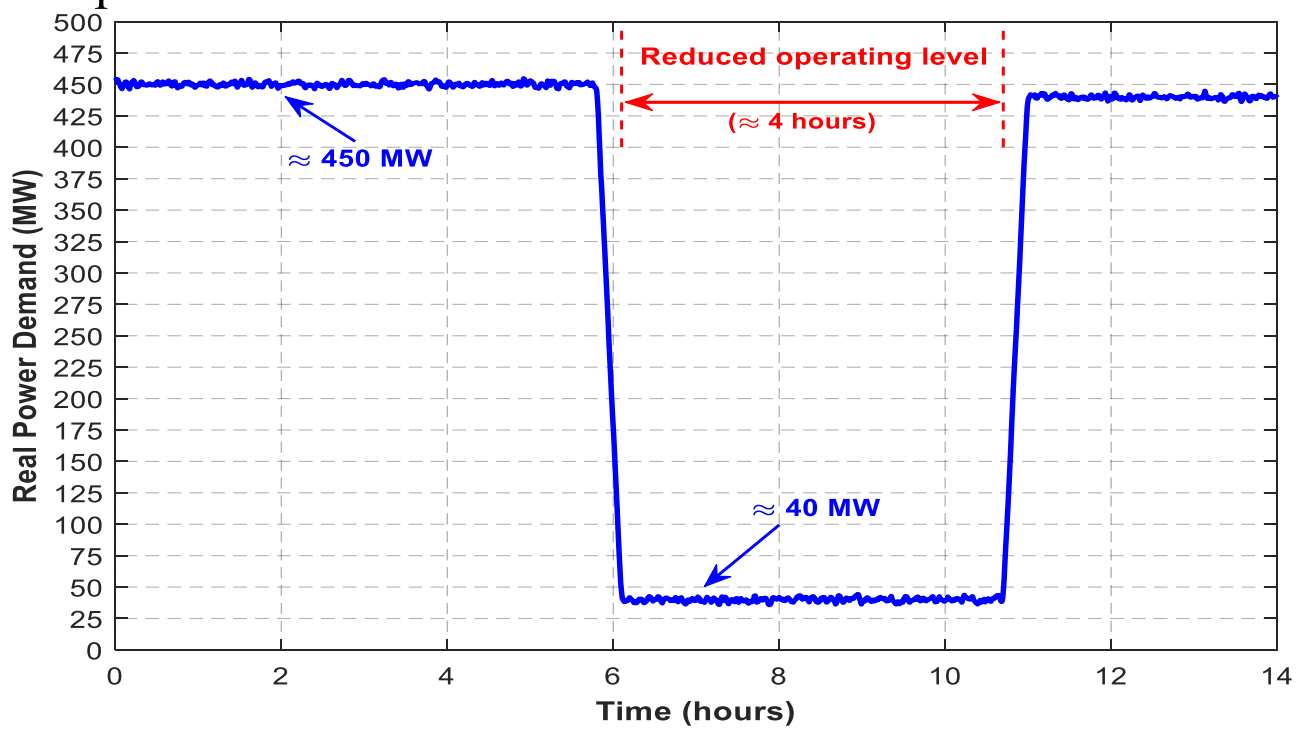

Fig. 1. Representative AI/data-center real-power demand ramp-down and ramp-up events [8].

This work was supported in part by the National Science Foundation of USA under Grants ECCS-2146615, CNS-2231523, and the eCAT Center industry members. Any opinions, findings, conclusions, or recommendations expressed in this material are those of the authors and do not necessarily reflect the views of the sponsors.

These characteristics reflect both the unprecedented scale of emerging AI-driven load growth and the operational fluctuations introduced by rapid workload-dependent ramping. Sudden multi-gigawatt load ramps from AI training events can mimic the system impacts of extreme-weather-induced outages. This could produce frequency excursions, voltage dips, or congestion patterns that strain conventional control reserves [9]. However, most existing resilience frameworks still focus almost exclusively on supply-side disruptions or assume static and predictable demand [10-13]. Few models explicitly represent the temporal uncertainty and clustering of these large flexible loads.

Therefore, this paper extends the disaster-resilience paradigm into a unified two-stage optimization framework capable of managing both physical disruptions and AI-driven demand uncertainty within a single mathematical structure. The framework introduces a new concept, Flexible Capacity Modules (FCMs), which unify generalized fast-response assets such as battery energy storage systems (BESS) [14], mobile diesel generators [15], and maybe future potential pumped-storage hydropower (PSH) sites into a single flexible dispatch category. These resources will be pre-allocated and dispatched through a Distributionally Robust Optimization (DRO) [16], model with Conditional Value-at-Risk (CVaR) [17], to hedge against uncertainty. By prepositioning FCMs near high-risk or high-volatility nodes (e.g., data center clusters), utilities can minimize the expected cost of unserved energy, curtailment, and imbalance during extreme events.

Classical optimization frameworks handle uncertainty primarily through Stochastic Programming (SP) [18], or Robust Optimization (RO) [19]. SP assumes known probability distributions and relies on extensive scenario sampling, whilst RO secures solutions against worst-case realizations without explicit probability information, which often yields overly conservative solutions. For instance, they strongly prefer to avoid low-probability, high-impact events such as widespread outages, even if this increases average operating cost [20]. To address these limitations, this work adopts a DRO framework augmented by a mean-CVaR formulation, providing a balance between robustness and risk-aware decision making [21, 22].

Motivated by these challenges, this paper develops a unified risk-aware resilience framework that simultaneously addresses AI-driven demand uncertainty and stochastic physical disruptions. The main contributions of this paper are threefold. First, a unified scenario framework is developed to simultaneously model AI-driven demand uncertainty, stochastic line outages, and repair-time variability. Second, a two-stage risk-averse DRO-MILP formulation is developed to coordinate the pre-positioning and post-event dispatch of FCMs, linking optimal pre-allocation and post-dispatch decisions to amplify the resilience of distribution networks. Third, the framework is validated on the IEEE 33-bus distribution system, demonstrating improvements in Energy Not Supplied (ENS), served-load recovery, and overall system resilience under severe operating conditions. The remainder of this paper is organized as follows. Section II presents the proposed two-stage DRO-MILP framework, Section III describes its mathematical formulation, Section IV discusses the case study and simulation results, and Section V concludes the paper.

## II. Proposed Two-Stage DRO-MILP Framework

This section presents the overall structure of the proposed two-stage DRO-Mixed Integer Linear Programming (MILP) framework for enhancing distribution-system resilience under both AI-driven demand uncertainty and stochastic physical disruptions. The framework coordinates the strategic pre-allocation and post-event dispatch of FCMs, including mobile BESS and mobile diesel generation (DG) resources. Although large AI data centers are typically connected at the transmission level, their operational impacts can be represented at the distribution level through equivalent load variations that capture feeder-level stress, congestion, and operational uncertainty.

### A. Framework Overview

The proposed framework extends traditional disaster-oriented resilience planning by simultaneously considering physical disruptions and AI-driven demand variability within a unified optimization structure. A two-stage risk-aware formulation is adopted, where first-stage decisions determine resource preparedness before uncertainty realization, while second-stage decisions coordinate operational response after the realization of AI demand fluctuations, outage events, and repair trajectories. The overall workflow of the proposed framework is illustrated in Fig. 2. In this work, AI-driven demand is represented through aggregate load trajectories without explicitly modeling workload migration, market-dispatch actions, or disruption-dependent operational responses.

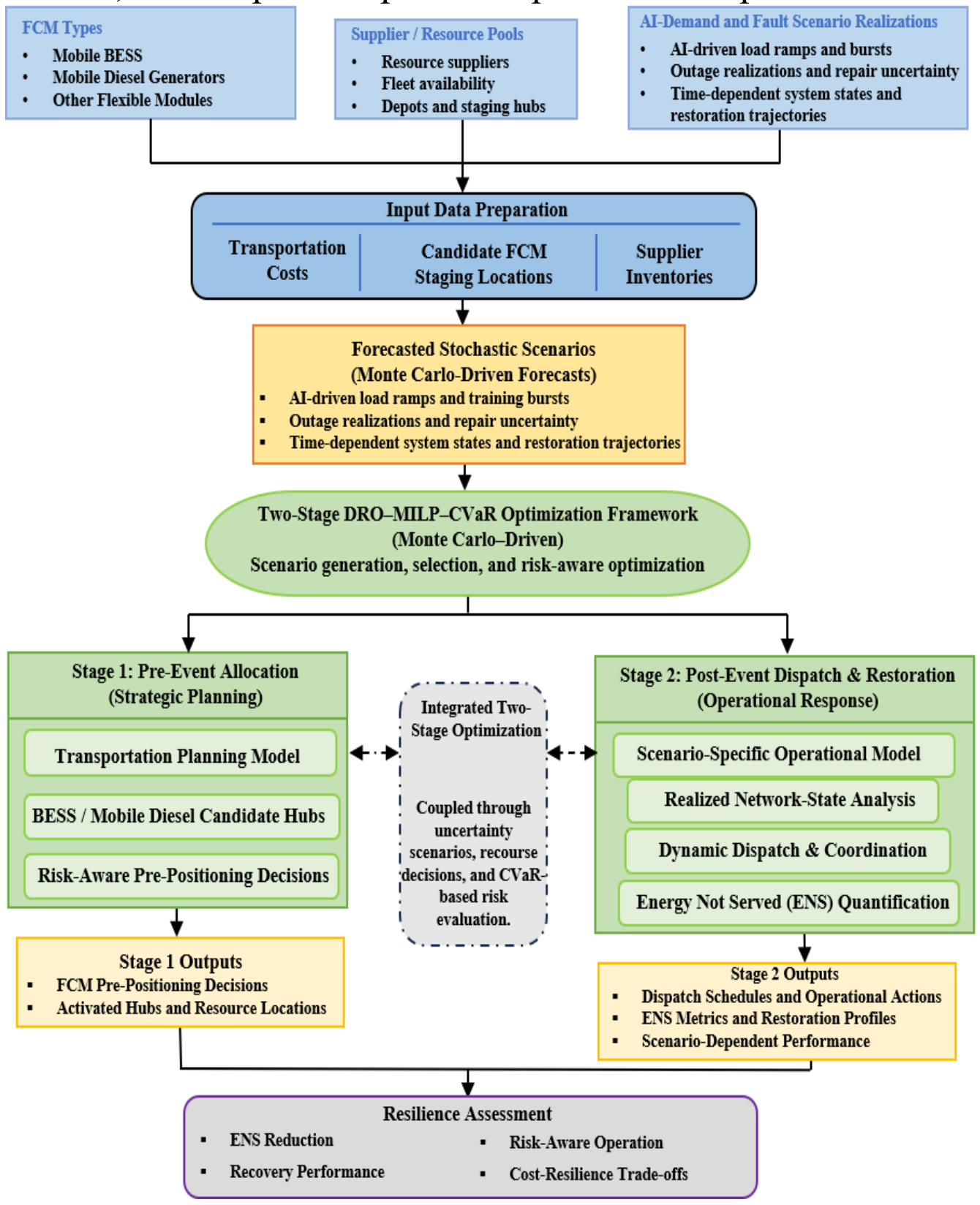

Fig. 2. Proposed two-stage optimization process flowchart for FCM allocation and dispatch under uncertain demand-shock scenarios.

### B. Stage 1: Pre-Event Allocation

Stage 1 determines the strategic pre-positioning of FCMs across candidate staging hubs located near critical or high-volatility load areas. These decisions are made before uncertainty realization and are therefore optimized using Monte Carlo (MC)-generated scenarios representing AI-driven demand variations, line outages, and repair-time uncertainty. The objective of Stage 1 is to minimize resource allocation and deployment costs while incorporating a mean–CVaR risk measure within a DRO framework. The resulting allocation decisions determine the quantity and location of FCMs available to be dispatched during the post-event.

### *C. Stage 2: Post-Event Dispatch and Restoration*

Following the realization of a scenario, Stage 2 determines the optimal dispatch and coordination of the pre-positioned FCMs. The operational model evaluates the realized network conditions, including AI-driven load variations, line availability, and computes corrective actions that support load recovery and system stabilization. Dispatch decisions are subject to network constraints, resource availability limits, and energy-state dynamics. The outputs of this stage include dispatch schedules, served-load recovery, restoration and ENS performance metrics.

### *D. Scenario* Generation and Risk Modeling

Uncertainty is represented through MC-generated scenarios that capture AI-driven demand fluctuations, stochastic line outages, and repair-time variability. To enhance robustness against uncertainty misspecification and extreme events, the proposed framework integrates a DRO formulation with a CVaR risk measure. This structure enables the model to balance expected system performance with protection against high-impact low-probability scenarios.

## III. Mathematical Formulation of the Two-Stage DRO-MILP Framework

This section presents the mathematical formulation of the proposed two-stage DRO-MILP framework. The model jointly determines pre-event allocation decisions for FCMs and post-event operational actions under uncertain AI-driven demand growth and stochastic disruptions. The formulation follows the general structure of resilience-oriented restoration models [23] while incorporating DRO and CVaR concepts to enhance protection against uncertainty and extreme events [17, 24].

### *A. Unified Scenario Representation*

The uncertainty realization associated with scenario $s$ is represented as in (1).

$$\omega_s = \left(P_{i,t,s}^{AI}, z_{e,t,s}^{line}, \tau_{e,s}^{rep}\right), \qquad \forall s \in S \tag{1}$$

where $P_{i,t,s}^{AI}$ denotes the AI-driven demand trajectory, $z_{e,t,s}^{line}$ represents the line-availability status, and $\tau_{e,s}^{rep}$ denotes the repair completion time of line $e$.

### *B. Two-Stage Optimization Structure*

The overall optimization problem is formulated as

$$\min_{x \in X}\left\{C^{(1)}(x) + \Phi(x)\right\} \tag{2}$$

Here $x$ denotes the first-stage decision vector, $C^{(1)}(x)$ represents the pre-event allocation cost, and $\Phi(x)$ denotes the risk-aware aggregation of the second-stage scenario-dependent recourse costs.

### *C. Stage-1 Pre-Event Allocation Model*

The Stage-1 planning cost is expressed as

$$C^{(1)}(x) = \sum_{d \in D} c_d\, z_d + \sum_{l \in L} \sum_{d \in D} b_l\, y_{l,d} \tag{3}$$

where $z_d$ is a binary variable indicating whether staging hub $d$ is activated, $y_{l,d}$ denotes the number of FCM units of type $l$ allocated to hub $d$, $c_d$ is the hub activation cost, and $b_l$ is the allocation cost coefficient associated with FCM type $l$.

The allocation decisions are subject to fleet availability constraints and hub-capacity constraints as in (4), and (5), respectively.

$$\sum_{d \in D} y_{l,d} \leq G_l, \forall l \in L \tag{4}$$

$$\sum_{l \in L} y_{l,d} \leq D_d^{\max} z_d, \forall d \in D \tag{5}$$

where $G_l$ denotes the total available fleet capacity of FCM type $l$, and $D_d^{\max}$ represents the maximum storage capacity of hub $d$.

### *D. Stage-2 Dispatch and Network Operation*

Following scenario realization, the operational recourse problem is given by (6).

$$Q(x, \omega_s) = \min_{y_s} C^{(2)}(y_s, \omega_s) \tag{6}$$

The recourse formulation in (6) follows the standard two-stage stochastic programming framework commonly adopted for post-disaster power-system restoration [25], where $Q(x, \omega_s)$ represents the minimum operational cost associated with scenario $s$.

Load shedding is computed as the difference between demand and served load according to (7), while the service limits are enforced through (8).

$$p_{i,t,s}^{shed} = P_{i,t,s}^{d} - p_{i,t,s}^{serve} \tag{7}$$

$$0 \leq p_{i,t,s}^{serve} \leq P_{i,t,s}^{d} \tag{8}$$

where $P_{i,t,s}^{d}$ is the total demand and $p_{i,t,s}^{serve}$ is the served load. The network feasibility is maintained through the line-capacity constraints in (9) and the nodal power-balance equations in (10).

$$-F_{ij}^{\max} z_{ij,t,s}^{line} \leq f_{ij,t,s} \leq F_{ij}^{\max} z_{ij,t,s}^{line} \tag{9}$$

$$\sum_{j:(j,i) \in E} f_{ji,t,s} - \sum_{j:(i,j) \in E} f_{ij,t,s} + p_{i,t,s}^{inj} = p_{i,t,s}^{serve} \tag{10}$$

Here $f_{ij,t,s}$ denotes line power flow and $p_{i,t,s}^{inj}$ represents local power injection.

### *E. FCM Operational Constraints*

The BESS state-of-charge (SOC) dynamics are modeled as

$$SOC_{b,t+1,s} = SOC_{b,t,s} + \eta_b^{ch} p_{b,t,s}^{ch} \Delta t - \frac{1}{\eta_b^{dis}} p_{b,t,s}^{dis} \Delta t \tag{11}$$

The SOC evolution is governed by (11), where $SOC_{b,t,s}$ denotes the BESS state and $\eta_b^{ch}$ and $\eta_b^{dis}$ represent the charging and discharging efficiencies. In addition to battery-based support, mobile DGs provide dispatchable backup power subject to output and fuel constraints given by (12) and (13) [25].

$$0 \leq p_{r,t,s}^{D} \leq P_r^{\max} \tag{12}$$

$$\sum_{t \in T} p_{r,t,s}^{D} \Delta t \leq F_r^{\max} \tag{13}$$

where $P_r^{\max}$ is the DG power limit and $F_r^{\max}$ is the available fuel-energy budget.

## F. Risk-Aware DRO-CVaR Formulation

To explicitly account for extreme high-cost scenarios, the CVaR risk measure is adopted following the formulation introduced in (14) [17].

$$CVaR_\alpha(Q) = \zeta + \frac{1}{1-\alpha}\sum_{s\in S} p_s\,\eta_s \tag{14}$$

subject to

$$\eta_s \geq Q_s - \zeta, \forall s \in S \tag{15}$$
$$\eta_s \geq 0, \forall s \in S \tag{16}$$

Here $Q_s$ denotes the recourse cost under scenario $s$, $\zeta$ is the Value-at-Risk threshold, and $\eta_s$ represents the excess loss above $\zeta$. To account for ambiguity in the scenario probabilities, the probability distribution is allowed to vary within the polyhedral ambiguity set defined in (17) [24].

$$\mathcal{P} = \left\{p: \sum_{s\in S} p_s = 1,\ p_s \geq 0,\ \hat{p}_s - \rho_s \leq p_s \leq \hat{p}_s + \rho_s\right\} \tag{17}$$

where $\hat{p}_s$ is the nominal scenario probability and $\rho_s$ is the ambiguity radius controlling allowable probability deviations. Combining the first-stage allocation cost, the DRO ambiguity set [24], and the CVaR risk measure yields the final risk-aware optimization model, represented as in (18).

$$\min_{x\in X}\left\{\underbrace{C^{(1)}(x)}_{\substack{Stage-1\\ Planning\ Cost}} + \max_{p\in\mathcal{P}}\left[(1-\lambda)\underbrace{\sum_{s\in S} p_s\,Q_s}_{\substack{Expected\ resource\ Cost\\ E\,[Q]}} + \lambda\underbrace{\left(\zeta + \frac{1}{1-\alpha}\sum_{s\in S} p_s\,\eta_s\right)}_{CVaR\ tail-risk\ term}\right]\right\} \tag{18}$$

where $\lambda \in [0,1]$ is the risk-aversion parameter that balances expected performance and tail-risk protection.

## G. Resilience Metric

System resilience is evaluated using the ENS metric, which quantifies the total curtailed energy throughout the restoration horizon and is computed as in (19).

$$ENS_s = \sum_{t\in T}\sum_{i\in N} p_{i,t,s}^{shed}\,\Delta t \tag{19}$$

# IV. Case Studies And Results

## A. Test System and Scenario Generation

The proposed framework is validated using the IEEE 33-bus radial distribution system operating at a nominal voltage level of 12.66 kV. To emulate emerging AI-driven load growth, selected buses are designated as AI/data-center demand nodes exhibiting stochastic demand behavior generated through MC-simulation. The AI-sensitive buses account for approximately 34% of the total system demand, representing a significant share of the feeder loading and enabling assessment of the resilience impacts associated with AI-driven demand uncertainty. DG units, BESS, and candidate FCM staging hubs are incorporated to support restoration during disruption events.

Uncertainty is represented through MC-scenario generation. Each scenario consists of a unique realization of AI-driven demand fluctuations, stochastic line outages, and repair trajectories. AI demand variability is modeled through training bursts, inference-load fluctuations, and stochastic ramping behavior. The resulting trajectories represent aggregate AI-data-center demand behavior and capture the large-scale ramping events reported for emerging hyperscale computing facilities. A total of 25 scenarios is considered in this study and are jointly evaluated within the proposed framework. The initial disturbance condition for the representative worst-case was Scenario 11, which is illustrated in Fig. 3.

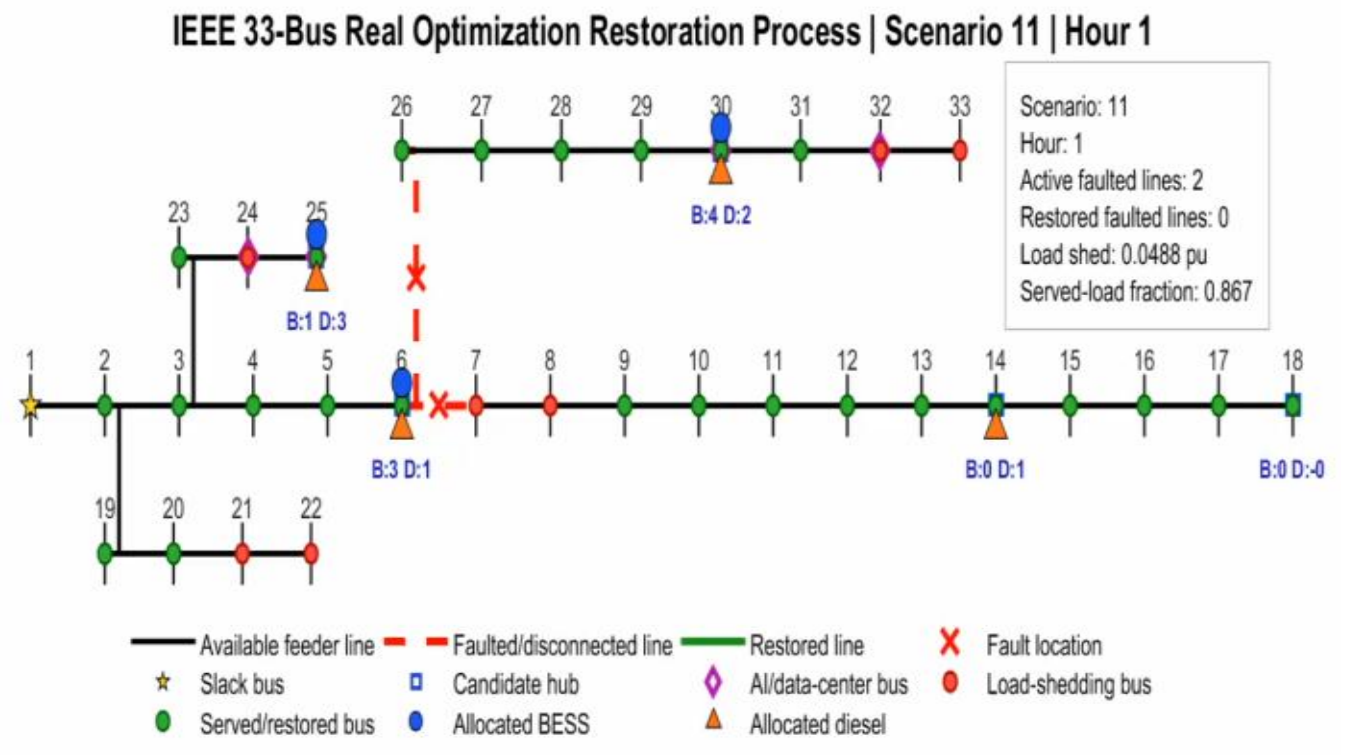

Fig. 3. Initial disturbance condition of the IEEE 33 bus system under Scenario 11 (Hour 1).

Fig. 4 presents the system configuration during the identified peak operational stress period at Hour 18 of Scenario 11. Compared with the initial disturbance stage shown in Fig. 3, this operating condition is characterized by elevated AI-driven demand levels and increased network loading. Although several feeder sections experience significant operational stress, the coordinated dispatch of FCMs maintains service continuity and preserves network feasibility, demonstrating the effectiveness of the proposed risk-aware restoration framework under severe operating conditions.

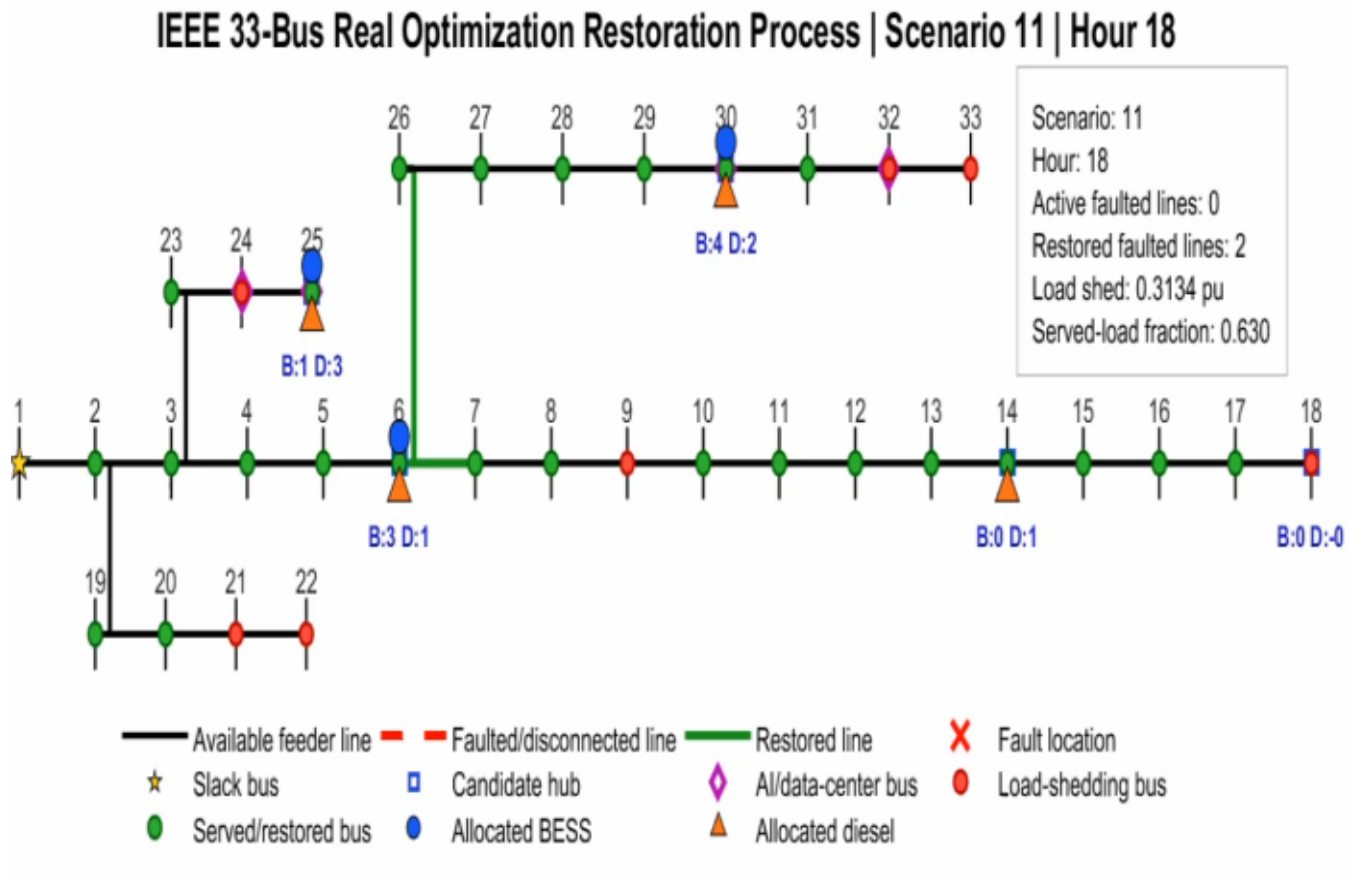

Fig. 4. IEEE 33 bus system during the peak operational stress stage under Scenario 11 (Hour 18).

## B. Optimization Performance

Table I summarizes the optimization and operational results obtained for the IEEE 33 bus system. The solver obtained a near-optimal integer-feasible solution with a relative optimality gap of 1.37%, confirming the computational tractability of the proposed DRO-MILP formulation. The results further indicate that resilience enhancement is achieved through relatively modest Stage-1 investment while maintaining acceptable voltage profiles and reducing exposure to high-impact operating conditions through the integrated DRO-CVaR framework.

Table I. Optimization and Operational Summary for IEEE System.

| Parameter | Value |
|---|---|
| Total optimization variables | 100,292 |
| Inequality constraints | 68,462 |
| Equality constraints | 43,200 |
| Integer variables | 6,015 |
| Relative optimality gap | 1.37% |
| Total objective value | $95,021.19 |
| Stage-1 planning cost | $1,580.00 |
| Expected recourse cost | $82,908.15 |
| CVaR contribution | $112,063.55 |
| Minimum voltage | 0.9500 pu |
| Maximum voltage | 1.0500 pu |

The Stage-1 allocation cost is significantly smaller than the expected recourse and risk-related costs, indicating that resilience enhancement is achieved through strategic resource positioning rather than excessive infrastructure deployment. The CVaR contribution of around $112,064 demonstrates the importance of protecting against severe high-impact scenarios. Consequently, the total objective value represents an optimized tradeoff among economic efficiency, restoration capability, and resilience robustness under AI-driven demand uncertainty and stochastic outages.

### *C. Resilience Performance and ENS Analysis*

ENS is adopted as the primary resilience-performance metric for evaluating restoration effectiveness under uncertain operating conditions. Fig. 5 presents the scenario-based ENS trajectories obtained from the generated MC-scenarios.

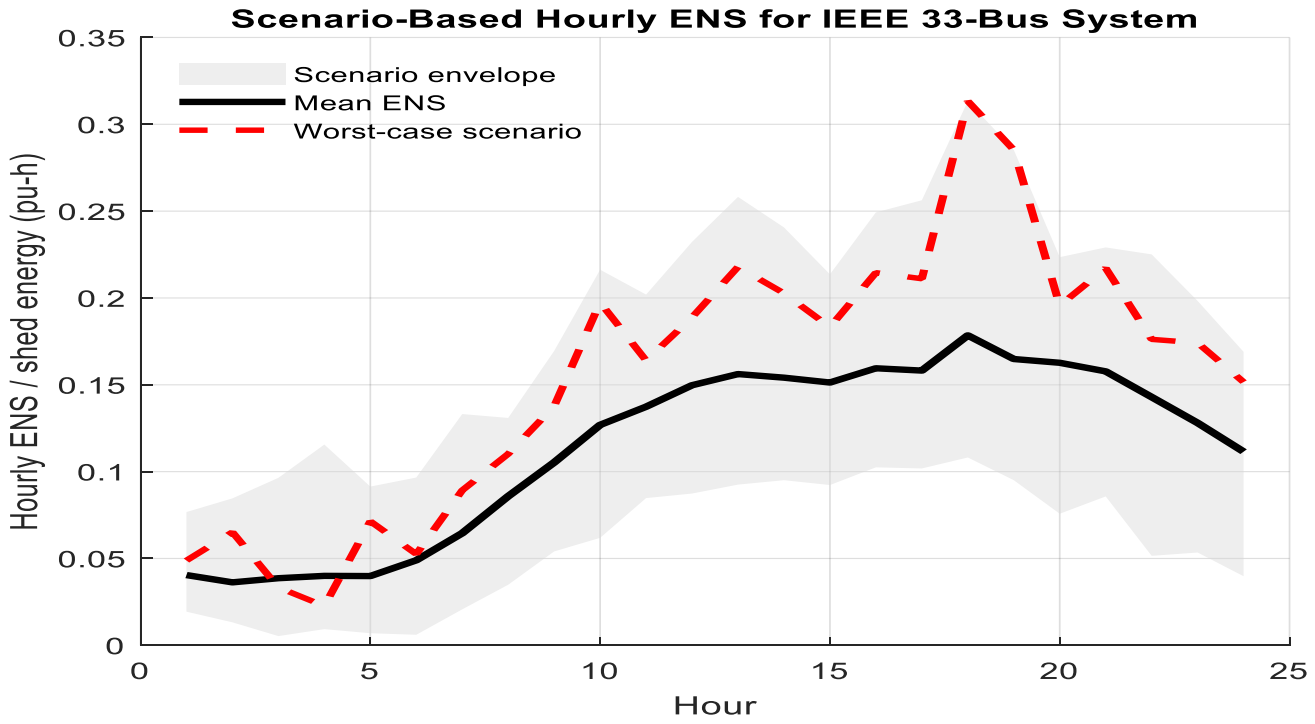

Fig. 5. Scenario-based hourly ENS trajectories for the system under stochastic outage and AI-driven demand conditions.

The results indicate that ENS gradually increases during periods of elevated operational stress caused by simultaneous outage conditions and AI-driven demand surges. The worst-case scenario reaches a maximum ENS value of 3.7241 pu-h, while the average ENS across all scenarios remains approximately 2.7393 pu-h. The relatively moderate spread between the minimum and maximum ENS values demonstrates that the proposed framework maintains resilience performance across a broad range of uncertainty realizations.

Furthermore, the integrated formulation reduces exposure to severe tail-risk events by explicitly accounting for both extreme operating conditions and ambiguity in scenario probabilities. Table II summarizes the statistical ENS characteristics and representative restoration metrics obtained across all generated MC-scenarios. The results provide insight into average system performance, scenario-to-scenario variability, and the effectiveness of the deployed restoration resources under uncertain operating conditions.

Table II. Scenario-Based ENS and Restoration Performance for the IEEE 33-Bus System.

| Metric | Value |
|---|---|
| Average ENS | 2.7393 pu-h |
| Minimum ENS | 1.9428 pu-h |
| Maximum ENS | 3.7241 pu-h |
| Worst-case scenario | Scenario 11 |
| Best-case scenario | Scenario 14 |
| Number of scenarios | 25 |
| Average diesel support | 8.8726 pu-h |

In addition to reducing ENS, the proposed framework successfully maintained voltage magnitudes within the prescribed operational limits throughout the restoration horizon. The minimum and maximum observed voltages were 0.95 pu and 1.05 pu, respectively, confirming that network feasibility and operational security were preserved despite stochastic feeder outages and AI-driven demand variability. Although localized voltage depressions were observed during periods of elevated operational stress, particularly near high-demand AI buses, the coordinated dispatch of mobile BESS and DG resources effectively mitigated severe voltage deviations. These results demonstrate that the proposed framework enhances both service restoration performance and voltage stability under uncertain operating conditions. The corresponding voltage profile under Scenario 11 is shown in Fig. 6.

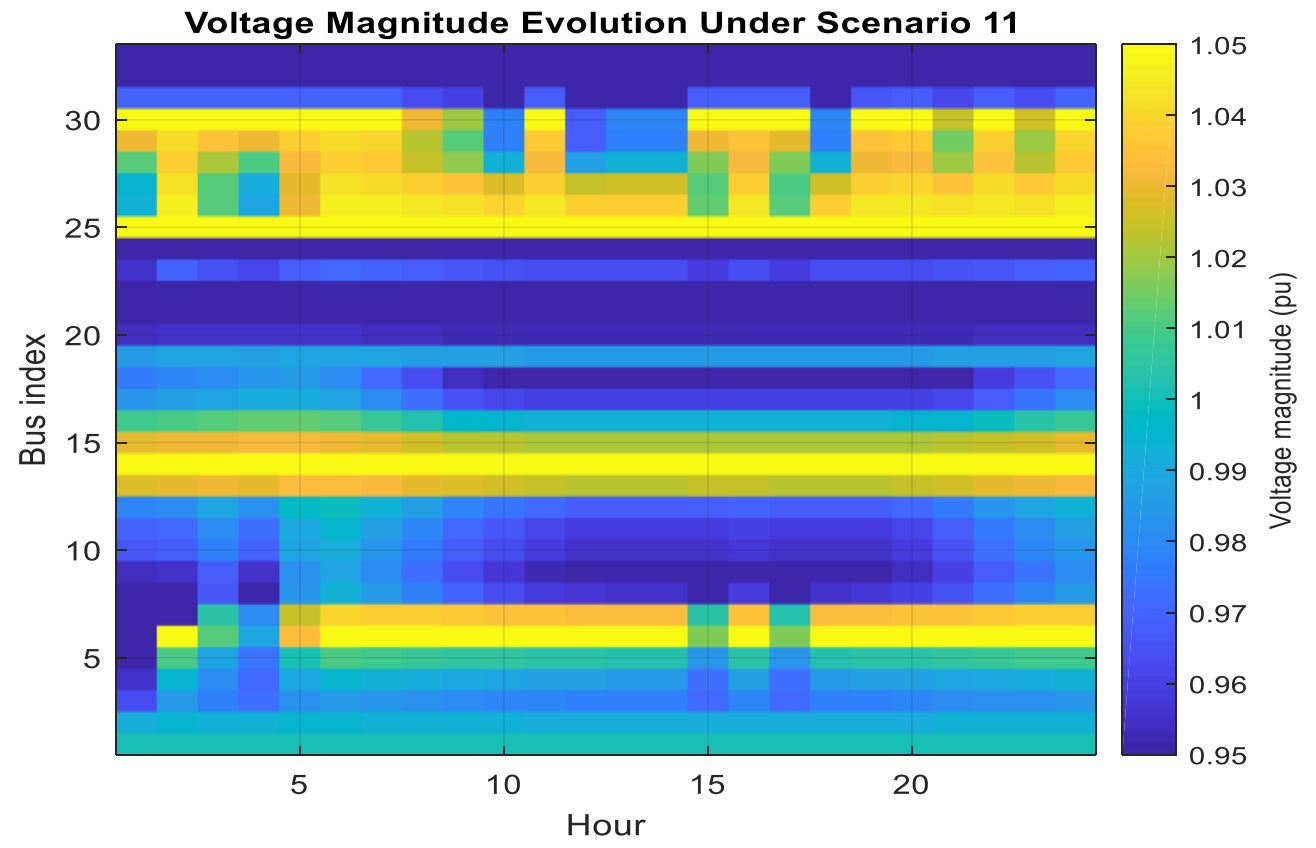

Fig. 6. Voltage magnitude profiles during disturbance and restoration periods for the system under Scenario 11.

### *D. FCM Allocation and Dispatch Behavior*

The Stage-1 allocation results demonstrate that the proposed framework selectively activates only the candidate hubs providing the highest resilience benefit. As summarized in Table III, Hubs 6, 14, 25, and 30 are activated, while Hub 18 remains inactive. Hub 30 receives the largest BESS allocation, whereas Hub 25 receives the largest diesel allocation, highlighting their strategic importance for supporting vulnerable feeder regions and AI-driven demand centers.

Table III. Optimal FCM Allocation Results

| Hub Location | BESS Units | Diesel Units | Hub Status |
|---|---|---|---|
| 6 | 3 | 1 | Active |
| 14 | 0 | 1 | Active |
| 18 | 0 | 0 | Inactive |
| 25 | 1 | 3 | Active |
| 30 | 4 | 2 | Active |

Fig. 7 illustrates the corresponding dispatch behavior for the representative worst-case scenario. In contrast, DGs provide sustained support throughout the restoration horizon. The complementary behavior of these resources demonstrates the value of combining fast-response energy storage with long-duration dispatchable generation. As a result, the proposed framework effectively enhances restoration capability while maintaining operational feasibility during severe operating conditions.

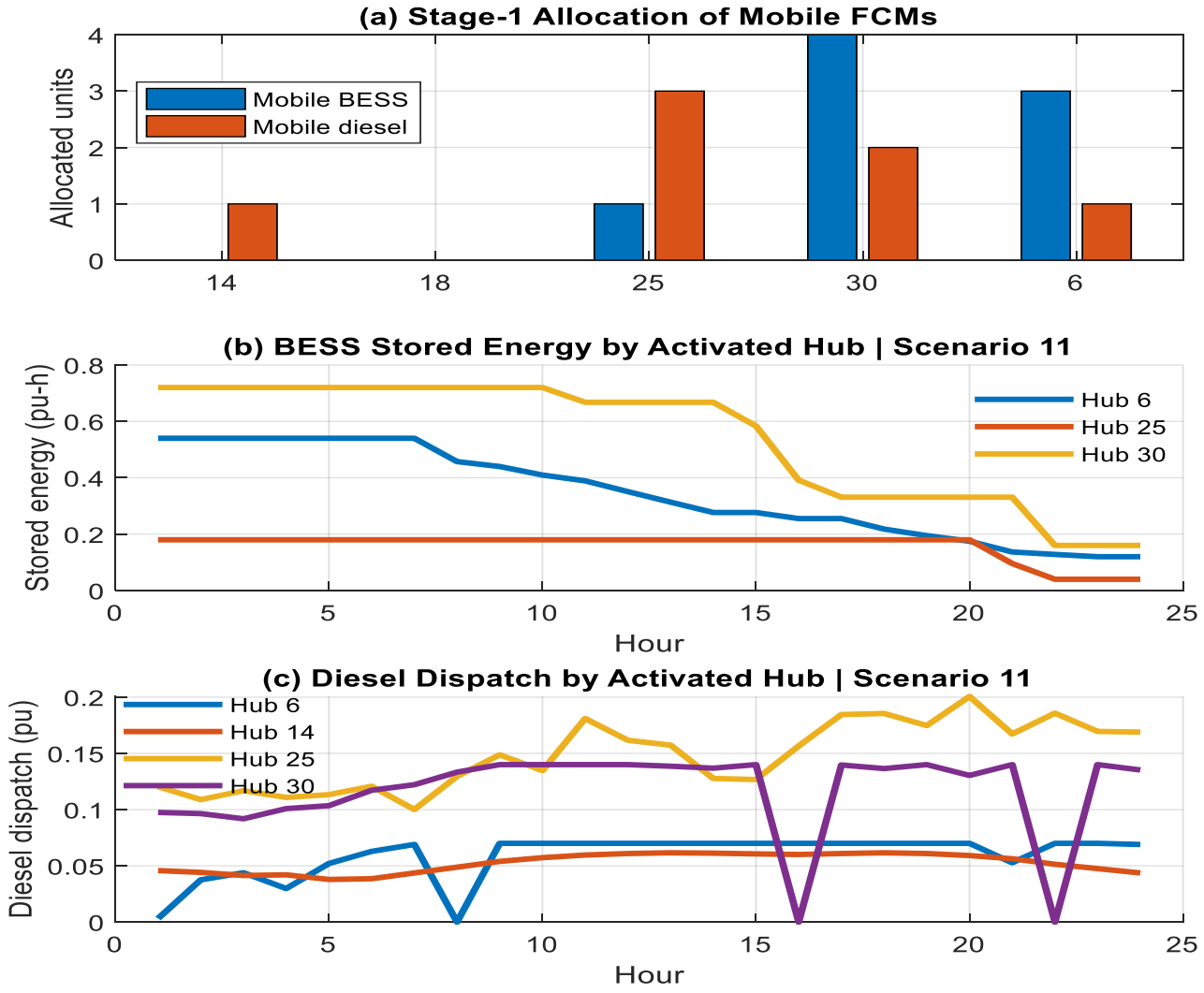

Fig. 7. Mobile-resource allocation and dispatch during restoration.

## V. Conclusion

A two-stage DRO-MILP framework has been presented for enhancing power-system resilience under AI-driven demand uncertainty and stochastic physical disruptions. The proposed model jointly optimizes pre-event allocation of FCMs and post-event operational restoration actions while explicitly accounting for uncertainty through a combined DRO-CVaR formulation. Simulation results obtained using the IEEE 33-bus distribution system demonstrate that the framework effectively reduces ENS, improves load-restoration capability, and strategically allocates mobile BESS and diesel resources under severe operating conditions. The results further demonstrate the effectiveness of coordinated FCM allocation and risk-aware operational decision-making in mitigating the impacts of stochastic outages and AI-driven demand variability.

Future work will investigate sensitivity analyses associated with uncertainty levels, risk-aversion parameters, and AI-driven load-growth scenarios, as well as the effects of workload flexibility, demand-response capabilities, and dynamic interactions between AI data centers and disruption events.